\documentclass[12pt]{article}
\usepackage{amsmath,latexsym}

\newcommand{\nor}{{\bf n}}
\newcommand{\tj}{{\bf j}}
\newcommand{\tk}{{\bf k}}
\newcommand{\tl}{{\bf l}}
\newcommand{\ti}{{\bf i}}
\newcommand{\oB}{\vert_{\partial\mathcal{M}}}

\newcommand{\indM}{\int_{\partial \mathcal{M}}}
\newcommand{\be}{\begin{equation}}
\newcommand{\ee}{\end{equation}}
\newcommand{\bea}{\begin{eqnarray}}
\newcommand{\rav}{\!\!\!&=&\!\!\!}
\newcommand{\ram}{\!\!\!&-&\!\!\!}

\newcommand{\eea}{\end{eqnarray}}
\newcommand{\pa}{\partial}
\renewcommand{\ll}{\left(}
\newcommand{\rr}{\right)}

\def\b{\beta}

\def\d{\delta}
\def\e{\epsilon}
\def\f{\phi}
\def\g{\gamma}

\def\j{\psi}

\def\l{\lambda}
\def\m{\mu}

\def\p{\pi}

\def\v{\varphi}

\def\.{\dot}
\def\2{{\frac 12}} \def\4{{1\over4}}
\newcommand{\ds}{/\hspace{-0.55em}\partial}

\begin{document}
\title{Consistent boundary conditions for supergravity}
\author{Peter van Nieuwenhuizen\\
{\it C.N.\ Yang Institute for Theoretical Physics}\\
{\it SUNY, Stony Brook, NY 11794-3840, USA}\\
\texttt{vannieu@insti.physics.sunysb.edu}
\\[5pt]
  Dmitri V.\ Vassilevich\thanks{Also at V.A.\ Fock Institute of Physics,
St.Petersburg University, 198904 St.Petersburg, Russia}\\
{\it Institut f\"ur Theoretische Physik, Universit\"{a}t Leipzig}\\
{\it Augustusplatz 10, D-04109 Leipzig, Germany}\\
\texttt{Dmitri.Vassilevich@itp.uni-leipzig.de}}
\date{YITP-SB-05/14 \\ LU-ITP 2005/018}
\maketitle
\begin{abstract}
We derive the complete orbit of boundary conditions for supergravity
models which is closed under the action of all local symmetries of these
models, and which eliminates spurious field equations on the boun\-dary. We
show that the
Gibbons-Hawking boundary conditions break local supersymmetry if one imposes
local boundary conditions on all fields. Nonlocal boundary conditions are not
ruled out. 
We extend our analysis to
BRST symmetry and to the
Hamiltonian formulation of these models.
\end{abstract}
\newpage
\section{Introduction}
Supergravity is supersymmetric general relativity. When it was
first constructed, as a field theory in $3+1$ dimensions with
$N=1$ gravitino,  partial 
integrations in the proof of local supersymmetry were performed
without taking boundary terms into consideration \cite{sugra}. 
However, it was
clear that in the presence of boundaries, local supersymmetry
(and other local symmetries such as local coordinate
invariance and local Lorentz symmetry\footnote{Local Lorentz symmetry is
an internal symmetry, but boundary conditions on spinors which are
not Lorentz invariant may lead to boundary conditions on the Lorentz
parameters.}) can only remain unbroken
if one imposes certain boundary conditions on the fields and
on the parameters. 

A natural arena to describe any theory with supersymmetry (called
henceforth susy) is superspace. One usually begins by defining
the integration over the anticommuting coordinates $\theta^\alpha$ as
ordinary Grassmann integration, $\int d\theta^\alpha$ with 
$\int d\theta^\alpha \theta^\alpha =1$ but $\int d\theta^\alpha 1=0$.
A more convenient way to identify the $x$-space component fields
contained in superfields is to use susy covariant derivatives
$D_\alpha = \frac{\partial}{\partial \theta^\alpha} +
i\sigma_{\alpha\dot\alpha}^\mu \bar \theta^{\dot\alpha}\partial_\mu $,
and to replace $\int dx d\theta^\alpha$ by $\int dx D_\alpha$.
As long as one may drop total $x$-derivatives, this makes no
difference, but it is clear that in the presence of boundaries
the results of these two approaches differ by boundary terms.
These boundary terms can be described by a boundary superspace
\cite{Lindstrom:2002mc}.

For ordinary gravity ($N=0$ supergravity), York \cite{York:1972sj}, 
and Gibbons and Hawking \cite{Gibbons:1976ue}, 
established long ago that one can cancel most of the boundary 
terms which one obtains if one varies the
metric in the Einstein-Hilbert action
by adding a boundary term which contains the extrinsic
curvature of the boundary.
A completely arbitrary variation of the metric in the sum of the 
Einstein-Hilbert bulk
action and the boundary action yields then the following result
\begin{equation}
\delta \mathcal{S}_{\rm EH} +\delta \mathcal{S}_{\rm bound}=
\int_{\mathcal{M}} G^{\mu\nu}\delta g_{\mu\nu} +
\indM (K^{\ti\tj} - g^{\ti\tj} K_{\tk}^{\tk})
\delta g_{\ti\tj} \,.\label{varEHGH}
\end{equation}
Here $ G^{\mu\nu}$ denotes the Einstein tensor, $\mathcal{M}$ is the
manifold with the usual measure $\sqrt{|\det g_{\mu\nu}|}$, 
$\partial \mathcal{M}$ its boundary with measure $\sqrt{|\det g_{\ti\tj}|}$\ , 
$K^{\ti\tj}$ is the extrinsic
curvature (see appendix \ref{ext-app}), and  indices $\mu$, $\nu$
refer to coordinates $x^\mu$ in the bulk, while indices $\ti$, $\tj$ refer to
coordinates $x^\ti$ in the boundary. We consider a boundary of 
dimension $D-1$ if spacetime has $D$
dimensions. Our results apply equally well to boundaries in time and
to boundaries in space.
We denote the coordinate which leads away from the surface
by $t$, even though it may be a spacelike coordinate.
Gibbons and Hawking proposed to impose the boundary condition that the
variations of the metric in the surface vanish
\begin{equation}
\delta g_{\ti\tj}\oB =0 .\label{dgij}
\end{equation}
We shall demonstrate that this violates local susy if one only admits local
boundary conditions (in particular no boundary conditions on curvature
components but only on the fields themselves).

We consider the case with zero cosmological constant only. For a non-zero
cosmological constant, the action on shell is infinite, but one can add
a boundary term to make it finite \cite{ccbc}. The cosmological constant
also brings an additional dimensional parameter in the bulk action
which can used to construct boundary actions. An example of such actions
is the ``boundary cosmological constant'' which appears in supergravity
theories with a bulk cosmological constant (see 
\cite{Luckock:1989jr,Liu:1998bu}).

The aim of the present article is to determine boundary conditions 
(called BC henceforth)
for pure supergravity theories which maintain all local symmetries.
Of course, not only the fields but also the gauge
parameters of these symmetry transformations must then be
restricted
on the boundary because gauge parameters become ghost fields in the
BRST formalism. 
It is important, that they are not over-restricted.
(For example, imposing both Dirichlet and Neumann BC on the same component
of a field or a parameter simultaneously
clearly over-restricts this field or parameter). 
In the BRST approach one requires 
the BRST invariance of the BC \cite{FraVil,Henneaux:1985kr}.
Several authors have already tackled many aspects of this problem,
in particular D'Eath, Luckock, Moss and Esposito 
\cite{D'Eath:1984sh,Luckock:1989jr,D'Eath:1996at,Esposito:1996kv} . 
Luckock and Moss 
\cite{Luckock:1989jr} have found an extra fermionic term in the
action on the boundary such that the whole action (bulk action plus
boundary action) is locally susy under variation with
a restricted susy parameter. They imposed the BC $\delta g_{\ti\tj}=0$,
but we shall pursue the question whether this condition is part of a
full set of BC which close under local susy (see below).
Even studies have appeared
with nonlinear sigma models on manifolds with boundaries
\cite{Luckock:1989jr,sigma}. 
We shall
instead consider pure supergravities, without extra fields. 
We consider both models with auxiliary fields and models without them.
Our aim is to
derive a complete and consistent set of 
BC. By this we mean the following. The total
set of BC should satisfy the following two
requirements
\begin{enumerate}
\item[(i)] it should not produce extra field equations on the boundary.
For example, we shall derive that one of the BC on
the fields themselves is $K^{\ti\tj}=0$, which evidently is the
alternative ``Neumann BC'' for gravity, instead of the ``Dirichlet BC''
$\delta g_{\ti\tj}=0$ for the variations in (\ref{varEHGH}).
($K^{\ti\tj}=-\frac 12 \partial_\nor g^{\ti\tj}$ in Gaussian coordinates,
where $\partial_\nor$ is the normal derivative). We impose these BC
on off-shell fields, even though they are derived from an analysis
of the field equations. Our aim is to use the total set of BC to define
the space of fields we consider both on- and off-shell. So we do not want
to begin with a set of BC for off-shell fields, and then later impose
separate BC for on-shell fields.
\item[(ii)] any rigid or local symmetry of the theory should transform any BC
into a linear combination of BC. For example, we shall require that
also the local susy variation of $K^{\ti\tj}$ vanishes, and this will yield
new BC. At the quantum level we shall
replace the set of all local symmetries by BRST symmetry, but the same 
requirement will be imposed. This leads to 
BC on the ghosts, as is well-known in string theory. It may seem 
contradictory to the non-expert that one gets different BC from 
field equations or local symmetries, because any local symmetry
variation can always be written as field equations times the local
symmetry variation of the fields. However, in order that variations 
cancel against each other one needs further partial integrations 
which lead to further BC. This is well-known among supergravity
practitioners.
\end{enumerate}

Before moving on, we should be clear about whether we impose BC on
background fields or on fluctuations about the background fields.
We only consider bosonic background fields.
We study first in section \ref{sec-lin} a trivial (flat space)
background in which case there are only fluctuating  fields, and
all BC refer to these fields. When one considers background fields
(for example, AdS space, or a black hole), we require that 
these background fields are susy, meaning that the susy 
transformation rules of the fermions vanish if one substitutes the
background fields into the right-hand sides of these transformation
rules. This leaves in general only rigid susy transformations
with susy parameters whose spacetime dependence is fixed.
Given such susy backgrounds, the transformation
rules reduce to rigid transformations for the 
fluctuation fields. However, even under this restricted class of 
rigid susy variations the classical action is not in general invariant
because boundary terms may remain. Then
BC for the background fields
like $K^{\ti\tj}=0$ can in general not be met, but only special boundaries
(for example boundaries spanned by geodesics) can achieve this. 
Alternatively, one can try to add boundary terms to the action such that the
BC for the background fields become satisfied. In section \ref{sec-kink}
we give a simple example of such boundary terms.
The set
of all fluctuating fields satisfying the complete set of BC
forms a linear vector
space, and the consistency of the BC means that symmetry transformations
never lead one out of this space.

Our strategy is as follows. We view the total set of BC as an orbit,
and we shall move freely forward and backward along this orbit, postponing
the solution of difficult constraints until we have obtained more information
from other constraints which are easier to solve. For example, one may
use information from local susy, in particular the closure of the gauge
algebra, to solve explicitly the equation $\delta K^{\ti\tj}=0$.
Since it still is true that supergravity is less well-known than general
relativity, we shall be very explicit and illustrate our results with 
a simple supergravity theory, $N=1$
supergravity in $2+1$ dimensions. (Real Majorana spinors for $N=1$ theories
require Minkowski spacetime instead of Euclidean space). Another strategy
we shall pursue is that we view those BC which are needed to preserve
local symmetries of the action as kinematical, in the sense that they
should not depend on the dynamics of a particular model. So, for
example, we may study pure gravity, to learn about the BC on the metric,
its variations, and the diffeomorphism parameters. Then we may use these
results in supergravity, for example requiring that the composite
parameter $\xi^\mu =\bar\epsilon_2 \gamma^\mu \epsilon_1$ in local
gauge algebra satisfies the same BC as $\xi^\mu$ in general relativity.
A very simple way to derive a subset of all BC is to consider 
a special case: free field
theories with rigid supersymmetries (section \ref{sec-lin}). 
In section \ref{sec-kink} we consider a model with a background: the susy kink.
As an amusing warming up exercise
for the full nonlinear supergravities we consider in section \ref{qm-sec}
a quantum mechanical model for supergravity where all nonlinearities
have a simple structure. In section \ref{sec-ham} we consider BC in a
Hamiltonian version of this model, and compare our results to those
obtained from the BRST formalism. In section \ref{sec21} we apply our
insights to the simple supergravity model mentioned above, and in section
\ref{sec-conc} we draw conclusions. 

In this paper we restrict our attention to local BC. Namely, for any
(multicomponent) field $\phi$ we define two complementary local
projectors on the boundary, $P_D$ and $P_N$, $P_D+P_N=1$ such that
$P_D\phi$ satisfies the Dirichlet BC $P_D\phi\oB=0$, and $P_N\phi$
satisfies the modified Neumann (Robin) BC 
$(\partial_\nor + \mathcal{S})P_N\phi\oB=0$, where $\mathcal{S}$
is a matrix-valued function on the boundary. Our analysis even allows for
$\mathcal{S}$ containing derivatives along the boundary up to a finite
order (as in the theory of open strings) although such derivative terms
do not appear in the particular models considered below. Such BC
with a non-derivative $\mathcal{S}$ are called mixed BC. More about
general properties of mixed BC one can learn from 
\cite{Gilkey:90,Luckock:1990xr,Branson:1999jz}. This restriction
to local BC looks rather natural. Indeed, the conditions we impose
on the fields at a given point of the boundary must not depend
on the value of the same field at distant points. Several authors
have used nevertheless nonlocal BC for supergravity 
\cite{Eath:1991sz,EKPbook,Esposito:1996kv}, but no closed locally
supersymmetric set (orbit) of such BC was found. In the present
paper we show that the Gibbons-Hawking BC (\ref{dgij}) cannot be
extended to a consistent locally susy orbit of {\em local} BC. 
This result may
indicate that one has to reconsider nonlocal BC in supergravity.

The supergravity community has not studied BC in detail in the past,
but the advent of string theory where BC play a crucial role may
also lead to further work on BC in supergravity from the same
perspective as in string theory\footnote{In several studies
of the AdS/CFT correspondence, BC on fields play a crucial role
\cite{ccbc}, but the invariance of these BC under local susy has not been
studied.}. 
Our article follows the same approach as an earlier article by
Lindstr\"{o}m, Rocek and one of us \cite{Lindstrom:2002mc}
on BC in superstring theory, and also in a paper
by the other author \cite{Vassilevich:2003xk} on the susy vortex.
Several of our results confirm results on BC in supergravity theories by
others, and we shall try to give references whenever this is possible.
However, we believe that the complete orbit of BC is new.
%%%%%%%
\section{Linearized supergravity with rigid susy}\label{sec-lin}
A simple and direct way to obtain (some) BC on fields and local
parameters is to consider linearized field theories with
rigid parameters. Thus we consider in this section linearized
supergravity in $3+1$ dimensions. The number of space-time dimensions
is not crucial. After obvious modifications the results of this section
will be valid, for example, also in $2+1$ dimensions. 
There are no background fields, so
all BC are on the fluctuating fields.
This analysis reveals the existence of two sets of
BC, one with $K^{\ti\tj}=0$, the other with $\delta g_{\ti\tj}=0$.
In section \ref{sec21} we shall consider the consequences of extending the
analysis to full nonlinear local susy, and we shall find that only 
the set with 
$K^{\ti\tj}=0$ is consistent in the sense described in the introduction.

It would seem natural to start with the linearized spin 2 and spin $3/2$
fields, but their analysis is rather complicated, and for that reason we
start at the other end, with the spin $0$ - spin $1/2$ system
and gather rather
efficiently information which will be of use for the spin 2 - spin $3/2$
system. Some of the statements regarding rigid susy of BC which
we derive below
are already known and collected in \cite{EKPbook}.

The action and susy transformations for a system consisting of a scalar
$S$, a pseudoscalar $P$ and a Majorana spinor $\lambda$ in four
dimensions read
\begin{eqnarray}
&&\mathcal{L} =
-\frac 12 (\partial_\mu S)^2 -\frac 12 (\partial_\mu P)^2
-\frac 12 \bar \lambda \gamma^\mu \partial_\mu \lambda \,,\label{free1200}\\
&&\delta_\epsilon S=\bar \epsilon \lambda ,\quad
\delta_\epsilon P=i\bar \epsilon \gamma_5 \lambda ,\quad
\delta_\epsilon \lambda =\left( \gamma^\mu \partial_\mu S +
i\gamma^\mu\partial_\mu P\gamma_5 \right) \epsilon .\nonumber
\end{eqnarray}
We are in Minkowski space.
In our conventions  $\gamma_5^2=1$,
$\gamma_5^\dag =\gamma_5$ and the $\gamma^j$ (with $j$ a space-like index)
are hermitian, while $\gamma^0$ is antihermitian. The symbol $\bar\lambda$
denotes the Dirac conjugate $\lambda^\dag i\gamma^0$ (which is equal to the 
Majorana conjugate $\lambda^TC$ for a Majorana spinor, with
$C$ the charge conjugation matrix). The action is real, and the susy
transformation rules preserve the reality properties of the fields.
The matrix 
$\gamma^\nor =\gamma_\mu n^\mu$ with $n^\mu$ the normal to the boundary
has no definite reality properties; in special cases it can be
hermitian or antihermitian,
but all formulas derived in the text hold for all cases,
essentially because $\lambda^\dag (\gamma^\nor )^\dag i\gamma^0 =
-\bar\lambda \gamma^\nor$. The Euler-Lagrange
variation of the fields in the action
(\ref{free1200}) leads to the boundary terms
$-\delta S \partial^\nor S -\delta P \partial^\nor P -
\frac 12 \bar \lambda \gamma^\nor \delta \lambda$, and following standard
arguments of string theory one concludes that there are four possibilities
in the spin 0 sector: Dirichlet conditions ($S\oB =0$,
$P\oB =0$) or Neumann conditions ($\partial^\nor S \oB =0$, 
$\partial^\nor P \oB =0$) for $S$ and $P$. 
One could add a boundary term ${\mathcal{L}}_{\partial\mathcal{M}}=
S\partial^\nor S$ (and a similar term for $P$). Then one would be
left with the boundary variation $S\partial^\nor \delta S$ instead
of $\delta S\partial^\nor S$ and the same BC would be obtained\footnote{
In ref. \cite{Schalm:2004qk} a boundary term 
$-\mu/2 \phi \partial_\nor \phi -\kappa/2 \phi^2$ is considered, 
and renormalization effects produce counterterms proportional to $\mu$.
A discontinuous field redefinition in their eq.\ (2.5) $\phi (x,y) \to
\phi (x,y) +\alpha \theta (y_0-y) \phi (x,y_0)$ with $y_0$ on the boundary, 
and a coupling  constant redefinition involving $\delta (0)$ absorbs $\mu$.}.
On the boundary one can at most restrict one half
of the spinor variables. We need BC for $\lambda$ without derivatives
to cancel $-\frac 12 \bar \lambda \gamma^\nor \delta \lambda$.
This leads to  one of the following
BC on $\lambda$: either $P_+\lambda \oB =0$, or $P_-\lambda \oB =0$, where
$P_+$ and $P_-$ are projection operators \cite{bagBC}
%\footnote{In Euclidean space
%they are given by $P_\pm =\frac 12 \left( 1\pm i \gamma^\nor \gamma_5 \right)$
%because if $P_+\lambda\oB =0$ one also requires $P_+\hat D \lambda \oB =0$
%where $\hat D=i\gamma^\mu \partial_\mu$ is hermitian. 
%(If $\hat D$ is hermitian, one
%can expand $\lambda$ over a complete set of eigenfunctions $\lambda_n$ each
%satisfying $P_+\lambda_n \oB =0$).}
\begin{equation}
P_\pm = \frac 12 (1\pm \gamma^\nor ).\label{Ppm}
\end{equation}
There is now no boundary term which can cancel (part of ) the boundary
variation $-\frac 12 \bar \lambda \gamma^\nor \delta \lambda$ since
$\bar\lambda \gamma^\nor \lambda$ vanishes.

The bulk field equations combined with these BC lead to a tower of
further BC involving even numbers of derivatives for the bosons and
powers of $P_+ \partial_\nor$ for the fermions. For example,
$S\oB =0$ and $\Box S=0$
leads to $\partial_\nor^2 S\oB =0$, $\partial_\nor^4 S\oB =0$
etc, while $P_-\lambda \oB =0$ is accompanied\footnote{Consider the field 
equation $\gamma^\mu \partial_\mu \lambda =0$. One finds by acting
with $P_+$ that $\gamma^\nor (P_+\partial_\nor \lambda )
+\gamma^\tj \partial_\tj (P_-\lambda)=0$. 
Hence if $P_-\lambda \oB =0$
(and thus also $\partial_\tj (P_-\lambda)\oB =0$) then 
$P_+\partial_\nor \lambda\oB =0$ on shell. 
In Euclidean space on has to choose different projection operators
$P^{(E)}_\pm =\frac 12 \left( 1\pm i \gamma^\nor \gamma_5 \right)$.
} by 
$P_+ \partial_\nor\lambda\oB =0$ etc. However, when we discuss the 
invariance of BC under symmetries of the action, we shall not
require that the fields satisfy their field equations.

In the presence of boundaries, one half of the susy is always violated.
Indeed, consider BC for the scalar. If one takes, for example,
$\delta S\oB =0$,
consistency requires that $\delta_\epsilon S\oB =\bar\epsilon\lambda\oB =0$.
If $P_\pm \lambda \oB =0$, one has to impose $\bar \epsilon P_\mp =0$
because $\bar\epsilon\lambda=\bar\epsilon (P_++P_-)\lambda$.
We suppose, that unbroken susy always corresponds to $\bar\epsilon P_+=0$,
which is equivalent to
\begin{equation}
P_-\epsilon =0 \,.\label{unbrosusy}
\end{equation}
The opposite choice ($P_+\epsilon =0$) leads to the equivalent results.
It is easy to see that there are two sets of BC which are invariant
under the susy transformations with the parameter restricted according
to (\ref{unbrosusy})
\begin{equation}
S \oB =0,\qquad \partial^\nor P\oB =0, \qquad P_-\lambda\oB =0
\label{set1-1}
\end{equation}
or 
\begin{equation} 
P \oB =0,\qquad \partial^\nor S \oB =0, \qquad P_+\lambda\oB =0\,.
\label{set2-1}
\end{equation}
Susy variations of these conditions produce again BC with extra 
$\partial_\nor$ derivatives, but now these BC are conditions for off-shell
fields in the action. For example, consistency of the BC in (\ref{set1-1})
leads to the further set $\partial_\nor^{2m} S\oB =0$,
$\partial_\nor^{2m-1}P\oB =0$, $P_- \partial_\nor^{2m} \lambda\oB =0$
and $P_+ \partial_\nor^{2m+1}\lambda \oB=0$ for $m=1,2,3\dots$

The contravariant index $\nor$ in $\partial^\nor S$ is defined by
$\partial_\mu (\delta S \partial^\mu S)
=\partial_\nor (\delta S \partial^\nor S)+\partial_\ti
(\delta S \partial^\ti S)$, where $\partial_\nor =n^\mu \partial_\mu$
and $\partial_\ti = \frac{\partial}{\partial x^\ti}$. 
It is lowered by the Minkowski
metric in the coordinate system with coordinates $(x^\nor ,x^\ti )$
along the normal and in the boundary.
 
Next we turn to the free spin $1$ - spin $1/2$ system
\begin{eqnarray}
&&\mathcal{L} =  -\frac 14 F_{\mu\nu}^2
-\frac 12 \bar \lambda \gamma^\mu \partial_\mu \lambda \,,\label{free112}\\
&&\delta_\epsilon A_\mu =\bar\epsilon \gamma_\mu \lambda ,\quad
\delta_\epsilon \lambda = - \frac 12 \gamma^{\mu\nu} F_{\mu\nu} \epsilon \,,
\nonumber  
\end{eqnarray}
where $\gamma^{\mu\nu}=\frac 12 (\gamma^\mu \gamma^\nu -\gamma^\nu\gamma^\mu)$.
In general, local BC for spin 1 fields can either be magnetic
($A_\tj\oB=0$) or electric
($F^{\tj\nor}\oB =0$) \cite{EKPbook}\footnote{On a spacelike boundary at fixed
time, the curvatures $F_{\tj\tk}$ and $F_{\tj 0}$ vanish
for these BC, respectively,
justifying the names magnetic and electric. On other boundaries we use
the same terminology.}. We split $F^{\nor\tj}\oB=0$ into the 
stronger set of BC $A_\nor\oB=0$
and $\partial_\nor A_\tj \oB=0$.
For both BC 
the boundary term $-\delta A_\tj F^{\tj\nor}$ produced by the Euler-Lagrange
variation vanishes. There are three boundary terms possible for $A_\mu$,
namely $A_\tj F^{\tj\nor}$, $A_\tj \partial^\nor A^\tj$ and
$A_\nor \partial^\nor A^\nor$. The first one leads to the boundary
variation $A_\tj \delta F^{\tj\nor}+\delta A_\tj F^{\tj\nor}$, 
and again the BC are the same as 
without this boundary term. The boundary terms $A_\tj \partial^\nor A^\tj$ and
$A_\nor \partial^\nor A^\nor$ are invariant under both sets of BC; in fact,
they vanish since we shall soon show that $A_\tj\oB=0$ implies
that also $\partial_\nor A_\nor \oB =0$.
So there is no boundary term for spin 1 either.
The other boundary term 
$-\frac 12 \bar\lambda \gamma^\nor\delta\lambda$ we discussed before.
However, only one of these BC is compatible
with susy for each of the BC for the spin $1/2$ field. 
As a result,
the following two sets are susy invariant
\begin{equation}
S \oB =0,\quad \partial^\nor P\oB =0, \quad P_-\lambda\oB =0,
\quad A_\nor \oB =0, \quad \partial_\nor A_\tj \oB=0
\label{set1-2}
\end{equation}
or 
\begin{equation}
P \oB =0,\quad \partial^\nor S \oB =0, \quad P_+\lambda\oB =0,
\quad A_\tj \oB =0\,.
\label{set2-2}
\end{equation}
Susy variations of these BC lead again to BC with additional $\partial_\nor$
derivatives as discussed above. For example, the susy variation of 
$P_+\lambda \oB =0$ in (\ref{set2-2}) leads to $F_{\ti\tj }\oB=0$ which
confirms $A_\tj \oB=0$. One would expect also $\partial_\nor A_\nor\oB=0$
in (\ref{set2-2}), and we shall indeed obtain this BC when we consider 
the spin $1$ - spin $3/2$ system.

We reach the spin $3/2$ level. The free spin 1 - spin $3/2$ system 
has the following action and rigid susy transformation rules 
\begin{eqnarray}
&&\mathcal{L} = -\frac 14 F_{\mu\nu}^2 
-\frac 12 \bar \psi_\mu \gamma^{\mu\rho\sigma} \partial_\rho \psi_\sigma ,
\label{free321}\\
&&\delta_\epsilon A_\mu =\bar \epsilon \psi_\mu ,\quad
\delta_\epsilon \psi_\mu = -\frac 12 F_{\mu\nu}\gamma^\nu \epsilon -
\frac i2 \tilde F_{\mu\nu}\gamma^\nu \gamma_5 \epsilon \,,\nonumber
\end{eqnarray}
where 
$\tilde F_{\mu\nu} =\frac 12 \varepsilon_{\mu\nu\rho\sigma}F^{\rho\sigma}$
with $\varepsilon^{0123}=1$. The fields $\psi_\mu$ are Majorana spinors.
To prove the susy invariance of the action
one needs the following identities $\varepsilon^{\sigma\tau\mu\rho}
\varepsilon_{\sigma\tau\alpha\beta} =-2(\delta_\alpha^\mu \delta_\beta^\rho
-\delta_\beta^\mu \delta_\alpha^\rho )$, $\gamma^{\mu\rho\sigma\tau}=
i\varepsilon^{\mu\rho\sigma\tau}\gamma_5$ and $\gamma^{\rho\sigma}\gamma_5=
-\frac i2 \varepsilon^{\rho\sigma\alpha\beta}\gamma_{\alpha\beta}$
\cite{Ferrara:1976fu}. 
The Euler-Lagrange variation of the action
yields the bosonic boundary term $-\delta A_\tj F^{\tj\nor}$ we discussed
above, and the spin $3/2$ boundary term 
$-\bar \psi_\ti \gamma^{\ti\nor\tj}\delta \psi_\tj$.
(Since the index $\rho$ in $\gamma^{\mu\rho\sigma}$ is along the normal,
the other indices $\mu=\ti$ and $\sigma =\tj$ lie in the boundary).
As in the case of spin $1/2$ there is no useful boundary term for
spin $3/2$ as $\bar\psi_\ti \gamma^{\ti\nor\tj} \psi_\tj$ vanishes.

Again we need BC on $\psi_\tj$ without $\partial_\nor$ derivatives.
It is clear that both for $P_+\psi_\tj\oB =0$ and $P_-\psi_\tj\oB =0$
this boundary term cancels. Since this time 
$\delta_\epsilon A_\ti =\bar \epsilon \psi_\ti$ instead of 
$\bar \epsilon \gamma_\ti \lambda$, the projection operators on $\lambda$
and $\psi_\ti$ must be opposite in order that no susy
breaking boundary terms occur. Similarly, $P_\pm \psi_\tj\oB =0$
requires $\delta_\epsilon P_\pm \psi_\tj\oB =0$. With the expression for
$\delta_\epsilon \psi_\tj$ given above, one finds the following conditions
\begin{eqnarray}
&&P_+\delta_\epsilon \psi_\tj\oB =
-\frac 12 F_{\tj\nor} \gamma^\nor (P_+\epsilon)\oB -
\frac i2 \tilde F_{\tj\tk} \gamma^\tk \gamma_5 (P_+\epsilon )\oB ,
\nonumber \\
&&P_-\delta_\epsilon \psi_\tj\oB =
-\frac 12 F_{\tj\tk} \gamma^\tk (P_+\epsilon)\oB -
\frac i2 \tilde F_{\tj\nor} \gamma^\nor \gamma_5 (P_+\epsilon )\oB .
\label{1500}
\end{eqnarray}
Thus $F_{\tj\nor}\oB =0$ if $P_+\psi_\tj\oB =0$, or $F_{\tj\tk}\oB =0$
if $P_-\psi_\tj\oB =0$. We recall that we split $F^{\nor\tj}\oB =0$ into
$\partial_\nor A_\tj\oB=0$ and $A_\nor\oB=0$.
Our two sets of BC increase as follows
\begin{eqnarray}
&&S \oB =0,\quad \partial^\nor P\oB =0, \quad P_-\lambda\oB =0,
\nonumber\\
&&A_\nor \oB =0,\quad \partial_\nor A_\tj \oB =0,\quad P_+\psi_\tj \oB =0 ,
\label{set1-3}
\end{eqnarray}
or
\begin{eqnarray}
&&P \oB =0,\quad \partial^\nor S \oB =0, \quad P_+\lambda\oB =0,
\nonumber\\
&&A_\tj \oB =0,\quad P_-\psi_\tj \oB =0.
\label{set2-3}
\end{eqnarray}

Susy variations yield further conditions. 
Consider first the set (\ref{set1-3}) and the susy variation 
$\delta_\epsilon F^{\nor\tj}\oB =\bar \epsilon 
(\partial^\nor \psi^\tj -\partial^\tj \psi^\nor)\oB =
\bar \epsilon 
(\partial^\nor P_- \psi^\tj -\partial^\tj P_-\psi^\nor)\oB$. For consistency
this expression should vanish. Since we consider local BC only, we like
to avoid nonlocal relations between $\psi^\tj$ and $\psi^\nor$ on the
boundary. Therefore, the two terms in the brackets above should vanish
separately,
\begin{equation}
P_-\partial^\nor \psi^\tj \oB =0,\qquad P_- \psi^\nor \oB =0 .\label{npsin}
\end{equation}
In the second set (\ref{set2-3})
we find the chain $P_-\partial_\nor\lambda\oB=0$,
$\partial_\nor F^{\ti\nor}\oB =0$ which implies 
$\partial^\nor A_\nor\oB=0$ (and $\partial_\nor^2 A_\ti \oB=0$), and
finally $P_-\partial^\nor \psi_\nor \oB =0$. So we recognize a pattern:
a given bosonic filed has BC with an odd number of normal indices
in one set while in the other set it has BC with an even number of normal
indices.

Now we are ready to analyze the free spin $3/2$ - spin 2 system.
Let us consider small fluctuations of the metric about flat Minkowski
background, $g_{\mu\nu}=\eta_{\mu\nu}+\kappa h_{\mu\nu}$. We expand the
action of $N=1$ supergravity in arbitrary dimensions, keeping at most
terms quadratic in fluctuations of $h_{\mu\nu}$ and 
in the gravitino $\psi_\mu$
\begin{equation}
\mathcal{L}^{\textrm{lin}}=\mathcal{L}^{\textrm{lin}}_{EH}
+\mathcal{L}^{\textrm{lin}}_\psi \,.\label{linact}
\end{equation}
The linearized Einstein-Hilbert gravity action reads
through second order in $h_{\mu\nu}$
\begin{eqnarray}
&&\mathcal{L}^{\textrm{lin}}_{EH}= \frac 1{2\kappa} (h^\mu_{,\mu}
- h_{,\mu}^{\ \ \mu} )
+\frac 12 \partial_\lambda \Omega^\lambda + \frac 12 \tilde{\mathcal{L}}^{(2)}
\label{linear}\\
&&\Omega_\lambda = h^{\mu\nu}\partial_\lambda h_{\mu\nu}
-h_{\lambda\nu} (h^\nu -h^{,\nu}) -h^{\mu\nu} \partial_{\mu} h_{\lambda\nu}\\
&&\tilde{\mathcal{L}}^{(2)}=-\frac 14 (\partial_\mu h_{\nu\sigma})^2
+\frac 12 (\partial_\lambda h_{\mu\nu} )(\partial^\mu h^{\lambda\nu})
-\frac 12 h^\mu h_{,\mu} +\frac 14 h_{,\mu}^2 \label{linEH}
\end{eqnarray}
Here $h_{\mu}\equiv \partial^\nu h_{\mu\nu}$, a comma denotes partial
differentiation, and $h\equiv h^\mu_{\ \,\mu}=h_{\mu\nu}\eta^{\mu\nu}$.

Of course one can always change $\tilde{ \mathcal{L}}^{(2)}$ by partial
integration, yielding corresponding changes in $\Omega_\lambda$.
The expression for $\tilde {\mathcal{L}}^{(2)}$
in (\ref{linEH})  corresponds  to the terms quadratic in $h_{\mu\nu}$ one
gets from minus the two $\Gamma \Gamma$ terms in  the Einstein-Hilbert
action.One often uses these latter two terms as action when one studies
canonical quantization or gravitational radiation, because in this form
there are no double derivatives of $h_{\mu\nu}$  in the action
\cite{textbooks}.
Note, however, that $\tilde{ \mathcal{L}}^{(2)}$ is not the
Fierz-Pauli action for free spin $2$ fields
\begin{equation}
{\mathcal{L}}^{(2)}_{\rm FP}=
-\frac 14 (\partial_\mu h_{\nu\sigma})^2
+\frac 12 (h_\mu)^2
-\frac 12 h_\mu h_{,\mu} +\frac 14 h_{,\mu}^2\,. \label{linFP}
\end{equation}
If one replaces $\tilde{ \mathcal{L}}^{(2)}$
 by $\mathcal{L}^{(2)}_{\rm FP}$  one finds that $\Omega_\lambda$ is
replaced by
$\Omega_{\lambda\ ({\rm FP})}
=h_{\mu\nu} \partial_\lambda h^{\mu\nu}
-h_{\lambda\nu}(\frac 32 h^\nu-h^{,\nu})$. We shall continue with 
(\ref{linear}) - (\ref{linEH}).

The linearized gravitino action was given above
\begin{equation}
{\mathcal{L}}^{\textrm{lin}}_\psi = 
-\frac 12 \bar \psi_\mu \gamma^{\mu\rho\sigma} 
\partial_\rho\psi_\sigma \label{linpsi}
\end{equation}
where for example $\gamma^{123}=\gamma^1\gamma^2\gamma^3$. 
Possible
boundary terms which may be added to the action (\ref{linear})
will be discussed later for each set of BC separately.

If one neglects all boundary terms the action
(\ref{linear}) is
invariant under the following rigid
susy transformations
\begin{eqnarray}
&&\delta_\epsilon h_{\mu\nu} 
=\frac 12 \left( \bar\epsilon \gamma_\mu \psi_\nu +
\bar \epsilon \gamma_\nu\psi_\mu \right), \quad
\delta_\epsilon \psi_\mu
 = \frac 14 \left( \omega_\mu^{\ \ mn} \right)_{\textrm{lin}}
\gamma_{mn}\epsilon \label{linsusy}\\
&&\left( \omega_{\mu mn} \right)_{\textrm{lin}} =\frac 12 \left(
\partial_n h_{\mu m} -\partial_m h_{\mu n} \right) \nonumber
\end{eqnarray}
This expression for the linearized spin connection\footnote{There is
also a term $-\frac 12 \partial_\mu (e_{mn}-e_{nm})$ in 
$\left( \omega_{\mu mn} \right)_{\textrm{lin}}$, but it does not contribute
to the susy variation of the action because it is a linearized
Lorentz transformation.}
is easily obtained from
(\ref{cone}), using 
$e_{\mu m}=\delta_{\mu m} +\frac \kappa{2} h_{\mu m}+\mathcal{O}(h^2)$.

We would like to extend the two sets of BC (\ref{set1-3}) and (\ref{set2-3})
to the gravitational field $h_{\mu\nu}$. Consider (\ref{set1-3}) first. 
We start with the requirement that the orbit of BC must be closed under the
rigid susy transformations (\ref{linsusy}).
The susy variation of
$h_{\tj\tk}$ on the boundary reads
\begin{equation}
\delta_\epsilon h_{\tj\tk} \oB = 
\frac 12 \left( \bar \epsilon \gamma_\tj P_- \psi_\tk
+\bar \epsilon \gamma_\tk P_-\psi_\tj \right) \oB =
\frac 12 \left( \bar \epsilon P_+ \gamma_\tj  \psi_\tk
+\bar \epsilon P_+\gamma_\tk \psi_\tj \right) \oB = 0 \label{1301}
\end{equation}
where we used (\ref{set1-3}) and (\ref{unbrosusy}). This equation
allows us to impose the BC $h_{\tj\tk}\oB =0$ where we
immediately recognize a linearized version of the Gibbons-Hawking
boundary condition (\ref{dgij}). Therefore, the first of our two sets
of BC reads
\begin{eqnarray}
&&S \oB =0,\quad \partial^\nor P\oB =0, \quad P_-\lambda\oB =0,
\nonumber\\
&&A_\nor\oB =0,\quad \partial_\nor A_\tj \oB =0, \quad
P_+\psi_\tj \oB =0 ,\quad h_{\tj\tk}\oB =0.
\label{set1-4}
\end{eqnarray}
This should be consistent with the BC $P_+\psi_\tj\oB =0$, whose susy variation
yield $(\omega_{\tj \tk\tl})_{\rm lin}\oB =0$. This is indeed consistent with
$h_{\tj\tk}\oB=0$. Consistency of the BC (\ref{npsin}) for $\psi_\nor$ yields
$\partial_\nor h_{\ti\nor}\oB=0$ and $h_{\nor\nor}\oB =0$, all with an even
number of $\nor$ indices.

Let us now now discuss which boundary terms should be added to the action
(\ref{linear}) to make this set of BC fully consistent. The Euler-Lagrange
variation of (\ref{linear}) yields a boundary term already at the 
{\em linear} order
\begin{equation}
-\frac 1{2\kappa} \int_{\partial\mathcal{M}} d^3x \partial_\nor 
\delta h_\ti^{\ \ti},\label{lvar}
\end{equation}
where we neglected the terms with $\partial_\ti \delta h^{\nor\ti}$ which are
total derivatives on the boundary. The occurrence of terms in the
variation which are linear in fields distinguishes gravity from other
field theories. Since we have already imposed the
Dirichlet BC on $h_{\ti\tj}$ we cannot impose also
the Neumann BC on the same components $h_{\ti\tj}$. Therefore, one
has to add a boundary term in order 
to cancel (\ref{lvar}). 
The only appropriate
boundary invariant is the trace of the extrinsic curvature integrated over
the boundary. By comparing this invariant to (\ref{lvar}) we fix the
coefficient in front of it and arrive at the York-Gibbons-Hawking
boundary term\footnote{Since the BC $h_{\tj\tk}\oB =0$ is not
preserved by general
coordinate transformations with $\xi^\tj$, the argument based on 3
covariance of the boundary term is not totally convincing. However,
in the quadratic order one an easily classify all possible boundary terms
containing two fields $h_{\mu\nu}$ and one derivative which can be 
added to the York-Gibbons-Hawking term. All such boundary terms either
vanish identically due to the BC which are already imposed, 
or their Euler-Lagrange variations
produce additional BC which overconstraint the system.}
\begin{equation}
{\mathcal{S}}_{\rm YGH}=\frac 1{\kappa^2} \int_{\partial{\mathcal{M}}}
d^3x \sqrt{|\det g_{\ti\tj}|} K_\ti^{\ \ti} \simeq
\frac 1{2\kappa} \int_{\partial\mathcal{M}} d^3x \partial_\nor 
\delta h_\ti^{\ \ti} + \mathcal{O} (h^2).\label{YGH}
\end{equation}
As a check one might prove that the boundary terms 
produced by the susy variation
and by the Euler-Lagrange variation of the action (\ref{linact}), supplemented
by the terms quadratic in $h_{\mu\nu}$ in the 
York-Gibbons-Hawking term (\ref{YGH}), vanish to next 
(quadratic) order as well if one uses (\ref{set1-4}) and 
$h_{\nor\nor}\oB=h_{\ti\nor,\nor}\oB=0$.
This calculation was done
in \cite{Luckock:1989jr} in the full non-linear theory and we shall not
repeat is here. We only note that the BC on $h_{\ti\tj}$,
$\partial_\nor h_{\ti\nor}$ and $h_{\nor\nor}$ require that
$\xi^\nor\oB=\partial_\nor \xi^\ti \oB =0$.
The authors of \cite{Luckock:1989jr} introduced a fermionic
boundary term which vanishes under the BC on the gravitino and therefore does 
not affect the proof. The paper \cite{Luckock:1989jr} did not obtain an
orbit of BC closed under the local susy transformations. (We shall show
in section \ref{sec21} that such an orbit with the 
Gibbons-Hawking BC on the metric fluctuations does not exist
for local BC).
%Instead, the paper \cite{Luckock:1989jr} proposed a set of boundary fields
%which are transformed into each other by the susy transformations.

Let us now turn to the other set of BC (\ref{set2-3}). Since 
$P_-\psi_\tj\oB =0$, for consistency
we also request $P_-\delta_\epsilon \psi_\tj\oB =0$.
This condition yields
\begin{equation}
0=P_-\delta_\epsilon \psi_\tj\oB = \frac 14 \left( \partial_\tk h_{\nor\tj}-
\partial_\nor h_{\tk\tj} \right) \gamma^{\nor\tk} P_+\epsilon \oB \,.
\label{1439}
\end{equation}
Therefore, we require
\begin{equation}
h_{\nor\tj}\oB =0\,,\qquad \partial_\nor h_{\tk\tj}\oB =0 .\label{1441}
\end{equation}
Now (\ref{lvar}) vanishes.
Next we compare (\ref{1441}) with the expression in (\ref{varK}) for the
extrinsic curvature. In the linearized
case (\ref{1441})  
implies that the extrinsic curvature vanishes
\begin{equation}
K_{\tj\tk}\oB =0 \label{curcond}
\end{equation}
Due to this BC the only boundary term we can add, namely 
the extrinsic curvature, vanishes together
with its susy variations.
Our second set of BC increases to
\begin{eqnarray}
&&P \oB =0,\quad \partial^\nor S \oB =0, \quad P_+\lambda\oB =0,
\nonumber\\
&&A_\tj \oB =0,\quad P_-\psi_\tj \oB =0, \quad h_{\nor\tj}\oB =0,
\quad \partial_\nor h_{\ti\tj}\oB =0.
\label{set2-4}
\end{eqnarray}
By closing this set with respect to other symmetry transformations
one arrives at further BC which we leave to the reader to derive.
They repeat the patterns we found before.
One can also obtain the remaining boundary conditions by taking the linearized
limit of the BC in full non-linear supergravity (see sec.\ \ref{sec21}).
This shows that
the set (\ref{set2-4}) should be supplemented 
by $\partial_\nor h_{\nor\nor}\oB =0$ in agreement with our rules for
the number of normal indices.
In particular we find
$h_\nor \oB = h_{,\nor}\oB =0$ in the second sector. By using this property
and (\ref{set2-4}),
it is easy to show that $\Omega_\nor$ vanishes on the boundary and
that one can integrate by parts in $\tilde{\mathcal{L}}^{(2)}$
without creating boundary terms. For the same reason one cannot write
a non-zero boundary term for $h_{\mu\nu}$. Indeed, any relevant boundary
term has the mass dimension one, i.e. it contains a single derivatives and,
consequently, an odd total number of vector indices. Since all tangential
(boundary) indices must be contracted in pairs, one has an odd number
of normal indices. All such terms vanish on the boundary.
%%%%%%%%
\section{Backgrounds with rigid susy}\label{sec-kink}
We next study consistent BC in a model with rigid susy and a
boundary term: the susy kink in 1+1 dimensions with the kink
soliton $\f_K(x)$ as background. The action reads
\bea
{\cal L} \rav  -\2(\pa_\m\f)^2+\2 F^2-\2\bar{\j}\ds\j-\2U'\bar{\j}\j +FU \\
U(\phi) \rav \sqrt{\frac{g}{2}}\ll \f^2-\frac{\m^2}{2g}\rr; \ \
\f=\f_K(x)+\eta(x,t);\ \ \pa_x\f_K+U(\f_K)=0
\eea
All fields are real.
The susy transformation rules for $\eta$, $F$ and
$\j$ with rigid parameter $\e$ read, using 
\begin{eqnarray}
&&\g^1= \ll \begin{array}{cc} 1 & 0 \\ 0
&-1 \end{array} \rr , \quad \g^0= \ll \begin{array}{cc} 0 & -1 \\ 1
& 0 \end{array} \rr ,\quad \j=\ll\begin{array}{c} \j_+ \\ \j_-
\end{array} \rr ,\quad \e=\ll\begin{array}{c} \e_+ \\ \e_-
\end{array} \rr
\nonumber\\
&&\delta\phi =\bar\epsilon \psi \,,\nonumber\\
&&\delta \psi = \gamma^\mu \partial_\mu \phi \epsilon +F\epsilon ,\\
&&\delta F=\bar\epsilon \gamma^\mu \partial_\mu \psi \,.\nonumber
\end{eqnarray}
Eliminating the auxiliary field $F$ by $F=-U$ yields a term $-\frac 12 U^2$
in the action, while the transformation rules become
\begin{eqnarray}
\delta\psi_+\rav\partial_x\phi\epsilon_+-\partial_t\phi
\epsilon_--U\epsilon_+ \,,\nonumber \\
\delta\psi_-\rav-\partial_x\phi\epsilon_-+\partial_t\phi
\epsilon_+-U\epsilon_- \, ,\label{kinksusy}\\
\d\eta \rav -i\e_+\j_-+i\e_-\j_+. \nonumber
\end{eqnarray}
The kink background $\f=\f_K$, $\j=0$ satisfies the field equation
$\pa_x\f_K+U(\f_K)=0$, and this background is clearly invariant
under susy transformations with $\e_-$. So all terms in (\ref{kinksusy})
are at least linear in quantum fluctuations.
Let us now study BC in
this model. We consider a boundary in space at fixed $x^1$.

From the field equations one finds the boundary term
\be
\int\limits_{-\infty}^{\infty} \left[ -\d\f\pa_x\f
-\2\bar{\j}\g^1\d\j\right] dt
\ee
For $\psi$ one finds as before $P_{\pm}\psi =0$, which becomes with
$P_{\pm}=\2 (1\pm\g^1)$ just $\j_{\pm}=0$. However, as it stands the
model cannot implement the Neumann BC $\partial_x \phi =0$ because
the background does not satisfy this condition. To remedy this, 
one can add a
boundary term $\int K(\f) dt$ \cite{Lindstrom:2002mc}. 
Then one finds the following
BC: $\d\eta (-\pa_x\f+K')=0$. Therefore, either $\eta\oB =0$ or
$(-\pa_x\f+K')\oB=0$. 
In order that the second BC (modified Neumann) holds
to zeroth order in $\eta$ one finds $K'=-U$. For the Dirichlet BC,
of course, no restrictions on $K$ follow but since $K=K(\phi_K)$
in that case, one may omit this boundary action altogether.
To linear order in
$\eta$ one finds from the field equations the following BC for the
fluctuations
\be\label{dm}
\eta_{| \pa M}=0 \ \mbox{ or } \ (\pa_x+U')\eta=0, \ \mbox{ and }\
\j_+=0 \ \mbox{ or }\  \j_-=0.
\ee

Rigid susy is preserved provided the boundary terms generated by
an $\e_-$ susy variation cancel. One
finds then to linear order in fluctuations two sets of
BC\footnote{To higher order in fluctuations one finds no new BC,
but rather these BC are modified by terms of higher order in
fluctuating fields.} which form a subset of~(\ref{dm})
\be
\left\{ \eta\oB =0, \ \ {\j_+}\oB =0 \right\} \mbox{
or } \ \left\{ (\pa_x+U')\eta\oB =0, \ \ \j_-\oB =0\right\}
\ee
The first set is closed under susy, but the second set leads to a
further BC
\be
\d(\e_-) (\pa_x+U')\eta \oB = (\pa_x+U')\j_+\oB=0
\ee
The latter BC, $(\pa_x+U')\j_+\oB=0$, transforms back into
$(\pa_x+U')\eta\oB$. So $\eta=0$ and $\j_+=0$ form a closed system,
as do of course $(\pa_x+U')\eta$ and $(\pa_x+U')\j_+=0$, but
we also found the BC $\j_- \oB =0$ in the second set, for reasons we
now explain.

In general one expects to need BC with $P_{\pm}\l=0$ {\em and}
$P_{\mp}(\pa_n\l+\dots)=0$ for fermions, and {\em either} $\eta=0$ or
$(\pa_n\eta+\dots)=0$ for bosons. We saw this 
happening in the action (\ref{free1200}), but due to the nontrivial 
background the Neumann conditions have now acquired extra terms.
The second set of BC indeed has
this form, but the first set misses a BC with $\pa_x\j_-$. The
nontrivial soliton background has already eliminated half of the
susy, and this seems to be the reason for the unexpected form of
the first set of BC. In ref \cite{Bordag:2002dg} it was shown that if one
imposes field equations one finds in the first set the ``missing
BC"
\be
(\pa_x-U'){\j_-}\oB =0\,,\label{1816}
\ee
but off-shell our approach does not lead to (\ref{1816}) as a BC.

%%%%%%
\section{Quantum mechanical supergravity}\label{qm-sec}
As a warming up exercise for theories with local susy
we now consider a simple model for supergravity where
all nonlinearities can easily be dealt with: the quantum mechanics of a bosonic
point particle $\varphi (t)$ and a one-component
fermionic point particle $\lambda (t)$
coupled to an external one-component gravitational field $h(t)$ and 
to an external one-component gravitino
field field $\psi (t)$. All fields are real.
(There do not, of course, exist gauge actions for
$h$ and $\psi$ in one dimension). This model is 
known to be locally supersymmetric
\cite{qmsugra}, and its BRST properties, both for the Lagrangian and the Hamiltonian
formulation, as well as its superspace formulation have recently been worked
out in \cite{vanNieuwenhuizen:2004td}. 
In none of these articles boundary terms 
have been discussed; that is the subject of this section. An interesting
aspect of this analysis is that one obtains boundary conditions in time.
In the next section we consider the Hamiltonian formulation of this model
and discuss possible boundary actions.

The classical action reads
\begin{equation}
L=\frac 12 (1-2h) \dot\varphi^2 +\frac i2 (1-2h) \lambda \dot\lambda 
-i\psi \dot\varphi \lambda \label{qmact}
\end{equation}
and is invariant under the following reparametrization and local susy
transformations
\begin{eqnarray}
&&\delta \varphi = \xi \dot\varphi +i\epsilon (1-2h)\lambda ,\label{vvar}\\
&&\delta \lambda = \xi \dot\lambda + \frac 12 \dot\xi \lambda 
-(1-2h) \dot\varphi \epsilon ,\label{vlam}\\
&&\delta h =\frac 12 \dot\xi +\xi \dot h -\dot\xi h -(1-2h) i\epsilon \psi ,
\label{vh}\\
&&\delta \psi = \xi \dot\psi -\frac 12 \dot\xi \psi + (1-2h) \left[
(1-2h) \dot\epsilon +\dot h \epsilon \right] \label{vpsi}
\end{eqnarray}
In this model
the susy Noether current 
$\dot\varphi\lambda$ in flat space varies into the current
$-(\dot\varphi \dot\varphi +i\lambda\dot\lambda)\epsilon$ which couples to $h$.
This is {\em not} the model one gets by putting the Dirac action 
$\frac i2 \lambda\dot\lambda$ in curved space, because even in curved 
space the Dirac action in one dimension remains $\frac i2 \lambda\dot\lambda$, 
without $h$ field.
However, one can rescale
 $\lambda$, $\psi$ and $\epsilon$, and then one finds the
action\footnote{\label{foot3}Both models are special cases of a 
one-parameter class of actions which are obtained from the Noether method.
The rescalings are $\psi =(1+2hx)^{1/2} \hat \psi$,
 $\epsilon =(1+2hx)^{1/2} \hat\epsilon$ and 
$(1+2hx)^{1/2} \lambda =\hat \lambda$.
For $x=0$ and $x=-1$ one finds actions in polynomial form, and $x=0$ gives 
(\ref{qmact}), while $x=-1$ yields the action (\ref{claction})
\cite{vanNieuwenhuizen:2004td}.} 
without coupling $ih\lambda\dot\lambda$. We shall use the latter model
for our BRST analysis , but continue for the time being with the former 
model; readers may of course interchange the models for either analysis.

The action is obtained by integrating $L$ over a finite time interval, 
and we study the BC at one of the two endpoints\footnote{This corresponds
to open string theory, but with BC in time, which have never been worked
out in string theory
as far as we know. For open strings one has two ghosts and antighosts
($c^+,\ c^-$ and $b_{++},\ b_{--}$) and then the BC become $c^+=c^-$
and $b_{++}=b_{--}$ instead of $c=0$ and $b=0$.}.
The field equations lead to the boundary terms 
\begin{equation}
\delta \varphi \left[ (1-2h) \dot\varphi -i\psi \lambda \right]
+\frac i2 (1-2h) \lambda \delta \lambda .
\end{equation}
These vanish provided\footnote{Another solution contains $h=1/2$; since in 
this case the whole action collapses, we do not consider this case.}
\begin{equation}
\lambda \oB =0 \label{qmbclam}
\end{equation}
and 
\begin{equation}
\varphi \oB =0\qquad\qquad \mbox{(Dirichlet)} \label{dirqm}
\end{equation}
or
\begin{equation}
\dot\varphi \oB = \partial_\nor \varphi \oB=0 \qquad 
\mbox{(Neumann)}.\label{neuqm}
\end{equation}
For a one-component fermion $\lambda$ one cannot, of course, define
$P_+$ or $P_-$.

From a general coordinate transformation one finds $\delta L=\frac d{dt}
(\xi L)$ (as usual, and one easily checks), so $\xi$ vanishes at the
boundary, $\xi\oB =0$. Making a local susy transformation yields the usual
boundary terms by partially integrating the kinetic terms of $\varphi$
and $\lambda$. This yields the following boundary terms
\begin{equation}
\dot\varphi i\epsilon (1-2h)\lambda 
-\frac i2 (1-2h)^2 \dot\varphi \epsilon \lambda .\label{1609}
\end{equation}
These terms vanish since we already know that $\lambda =0$ at the
boundary. So the action is Einstein (general coordinate) and local susy
invariant, but there is of course no local Lorentz invariance to be dealt
with in this model.

Consistency of $\lambda =0$ and either $\varphi =0$
or $\dot \varphi =0$ requires that also 
symmetry transformations of these constraints vanish at the boundary. 
This is the case for $\xi$ transformations if $\xi$ vanish at the boundary
\begin{equation}
\xi\oB =0.
\end{equation}
For example $\delta_\xi \dot \varphi =\xi \partial_t^2 \varphi +
\dot\xi \dot \varphi =0$ when $\xi =0$ and $\dot\varphi =0$.
For local susy we find from $\lambda =0$
\begin{equation}
0=\delta_\epsilon \lambda \oB =(1-2h) \dot\varphi \epsilon \oB \label{1617}
\end{equation}
Consequently, $\epsilon\oB =0$ if $\varphi\oB=0$. If $\varphi\oB =0$ we obtain from
$\delta_\epsilon \varphi =
i\epsilon (1-2h)\lambda$ so we find no new BC, but if $\dot\varphi =0$ we find
from $\delta_\epsilon \dot\varphi\oB =0$ that also in this case $\epsilon$
vanishes at the boundary
\begin{equation}
\epsilon\oB =0 .\label{1622}
\end{equation}

There are no auxiliary fields needed in this model, 
and the local gauge algebra indeed
closes. One finds
\begin{equation}
[\delta_s (\epsilon_2),\delta_s (\epsilon_1) ]=
\delta_s (\tilde\epsilon = -2i\epsilon_2\epsilon_1 \psi )
+\delta_g (\tilde \xi =(1-2h)2i\epsilon_2\epsilon_1).\label{1636}
\end{equation}
Consistency requires that $\tilde\xi$ and $\tilde\epsilon$ vanish
at the boundary as they clearly do. Other commutators read
\begin{eqnarray}
&&[\delta_s(\epsilon),\delta_g (\xi)] 
=\delta_s (\tilde\epsilon =\xi\dot\epsilon ),\nonumber\\
&&[\delta_g (\xi_2),\delta_g(\xi_1)]
=\delta_g ( \tilde\xi =\xi_1\dot\xi_2 -\xi_2\dot\xi_1) \nonumber
\end{eqnarray}
and also this time $\tilde\xi$ and $\tilde\epsilon$ vanish at the boundary.

Consider next the BRST symmetry. We start from the classical
action\footnote{We mentioned this action in footnote \ref{foot3}.
It is this form of the action which can be straightforwardly written
in superspace \cite{vanNieuwenhuizen:2004td}.}
\begin{equation}
L_{\rm cl}=\frac 12 \dot\varphi^2 +\frac i2 \lambda\dot\lambda -
h \dot\varphi^2 -i\psi \dot\varphi \lambda \label{claction}
\end{equation}
and add the nonderivative gauge fixing term one uses in string theory
\begin{equation}
L_{\rm fix}=dh +\Delta \psi ,\label{fixact}
\end{equation}
where $d$ and $\Delta$ are the BRST auxiliary fields. 

The BRST rules are 
\begin{eqnarray}
&&\delta_B \varphi = [\dot\varphi c -\lambda\gamma ] \Lambda ,\nonumber\\ 
&&\delta_B \lambda = [\dot\lambda c +i(1-2h) \dot\varphi \gamma +
\psi\lambda\gamma ]\Lambda ,\nonumber \\
&&\delta_B h = [\frac 12 (1-2h) \dot c + \dot h c +(1-2h)\psi\gamma ] \Lambda ,
\nonumber \\
&&\delta_B \psi = [\dot\psi -i (1-2h) \dot \gamma ]\Lambda ,\label{BRSTrules}
\end{eqnarray}
where $\Lambda$ is the constant anticommuting imaginary BRST parameter,
the real $c$ is the coordinate ghost ($\xi =\xi \Lambda$) and the
real $\gamma$ is the susy ghost ($\epsilon =-i\gamma\Lambda$). The ghost 
action becomes
\begin{eqnarray}
L_{\rm ghost} &=& b\left[ \frac 12 (1-2h)\dot c +\dot h c +
(1-2h) \psi \gamma \right] \nonumber \\
&&+\beta \left[ -i (1-2h)\dot \gamma +\dot\psi c \right] \label{1712}
\end{eqnarray}
where the antihermitian $b$ is the coordinate antighost and 
the antihermitian $\beta$ the susy antighost. The transformation rules
of the ghosts follow from the closure of the local gauge algebra
(or from the invariance of the action)
\begin{eqnarray}
&&\delta_B c = [-c\dot c +i(1-2h)\gamma\gamma ]\Lambda \nonumber \\
&&\delta_B \gamma = [c\dot \gamma  +\psi \gamma\gamma ]\Lambda 
\label{1542}
\end{eqnarray}
and as usual the antighosts and the auxiliary fields form contactable
pairs
\begin{eqnarray}
\delta_B b=\Lambda d,\qquad \delta_B d=0\nonumber\\
\delta_B \beta=\Lambda \Delta,\qquad \delta_B \Delta =0. \label{1543}
\end{eqnarray}

From the field equations we obtain again the BC in (\ref{qmbclam}) -
(\ref{neuqm}), and further
\begin{equation}
(b\delta c)\oB=0,\quad (\delta hbc)\oB=0,
\quad (\beta\delta \gamma)\oB =0,\quad
(\beta \delta \psi c)\oB =0.\label{extra}
\end{equation}
Before solving these, we consider other points on the orbit of BC.
Let us check that BRST symmetry preserves the boundary conditions. 
From $\delta_B \lambda\oB =0$ one obtains
\begin{equation}
c\oB =0,\qquad (\dot\varphi \gamma )\oB =0 \end{equation}
From $\delta_B \varphi \oB =0$ one finds $c\oB =0$ and 
$\lambda \gamma \oB =0$, both of which are satisfied. Since 
$\dot\varphi \oB \ne 0$ in this case, we also find that
\begin{equation}
\gamma\oB =0 \end{equation}
On the other hand, $\dot\varphi\oB =0$ implies that
$\dot\lambda \gamma =0$, and since $\dot\lambda \oB \ne 0$, we conclude
that also in this case the susy ghost vanishes at the boundary.

Hence, all ghosts vanish at the boundary. This is due to the algebraic gauge
choice $h=\psi=0$; for de Donder type of gauge for $h$ or a gauge choice
$\psi \frac d{dt} \psi$ for $\psi$ one would get different results. 
Also the BC for the (anti)ghosts in (\ref{extra}) are then satisfied.
For consistency
the BC $c\oB =\gamma\oB =0$ requires that also 
$\delta_Bc\oB =\delta_B\gamma\oB =0$. This is indeed the case
as one checks from (\ref{1542}), 
and hence we
conclude that the total consistent set of BC consists of
\begin{eqnarray}
&&\lambda\oB =0,\quad \varphi\oB =0 \ \mbox{or}\ \dot\varphi\oB =0,\quad
\nonumber \\
&&c\oB =0,\quad \gamma\oB=0,\quad \xi\oB=0,\quad \epsilon\oB =0 
\end{eqnarray}
There are no BC on $h$, $\psi$, $b$, $\beta$ in this quantum mechanical
model with one-component fields. We consider possible boundary terms in the 
next section.
%%%%%%%
\section{Hamiltonian boundary conditions}\label{sec-ham}
So far we have been discussing models in the Lagrange formalism.
In the Hamiltonian formalism the issue of BC is simpler because
the action is of the form $L=\dot{Q}P-H$ where $H=H(Q,P)$ does not
contain any derivatives. BC enter then the path
integral as conditions on the states at initial and final times
\cite{Henneaux:1985kr}.Conditions involving time derivatives,
of the form $\partial_\nor \varphi =0$, become now conditions on
momenta, hence one needs only to specify the values of (half of the)
fields and momenta at the boundary.
If these states are physical states, they should be annihilated by
the BRST charge $Q$. This raises the question whether $Q=0$ is
equivalent to the BC one gets from our program. Our BC are, of course,
off-shell, whereas the ones from $Q=0$ are on-shell.
We use the quantum
mechanical model (\ref{claction})
again to study these issues in a concrete way.

The action in Hamiltonian form is given by
\begin{equation}\label{acti}
L={\dot \varphi}p +{\dot \lambda}\pi_{\lambda}+{\dot G}p_G + {\dot
\Psi}\pi_{\Psi} +{\dot c}\pi_c +{\dot b}p_b+{\dot
\gamma}p_{\gamma} + {\dot \beta}\pi_{\beta}+\{Q_H, \psi_g\}.
\end{equation}
where $Q_H$ is the nilpotent quantum BRST charge in the
Hamiltonian formalism, which has in our case the form
\be\label{pdv4}
Q_H=\frac12c p^2 - i\gamma p(\pi_{\lambda} -\frac{i}{2}\lambda)
+p_bp_G+\pi_{\beta}\pi_{\Psi} - i\pi_c\gamma\gamma,
\ee
In a general Hamiltonian framework the action has the form
\be
L = \dot{q}^i p_i - H + \{ Q_H, \psi_g \}, \label{Ham}
\ee
but the quantum Hamiltonian $H$ which commutes with $Q_H$ vanishes
in our case. The transformation rules which leave the classical
action invariant up to boundary terms are as follows. The
diffeomorphisms are generated by $\2 p^2$. This yields
\bea\label{pdv}
&& \delta\varphi={\hat\xi}p, \quad \delta p=0, \quad
\delta\lambda=0,\quad
\delta\pi_\l=0, \nonumber\\
&& \delta\Psi=0,\quad \delta (1+2H)= \frac{d}{dt}{\hat \xi},\quad
{\hat \xi} = (1+2H)\xi. \eea  
The classical gauge fields are
$G=(1+2H)$ and $\Psi$ and they transform in general as
\begin{equation}\label{AAA}
 \delta h^A = \frac{d}{dt} \epsilon^A + f^A_{\ BC} h_{\mu}^B \epsilon^C,
\end{equation}
where $f^A_{\ BC}$ are the structure functions of the local gauge
algebra. The local susy
transformations are generated by $ip(\pi_\lambda -\frac i2 \lambda)$.
One finds using Dirac brackets,
\bea
&& \delta\varphi =- \epsilon (\pi_{\lambda} -\frac{i}{2}\lambda),\
\ \ \delta p = 0, \ \ \ \delta\lambda=-p\epsilon,\ \ \
\delta\pi_{\lambda}=\frac{i}{2}p\epsilon. \nonumber \\
&&\delta\Psi = {\dot
\epsilon },\qquad \delta (1+2H)= -2i\epsilon\Psi.\label{pdv2}
\eea

We choose as gauge fermion
\be
\j_g=-i G \p_c-\j p_\g
\ee
and find for the gauge-artefacts
\be
\{Q_H,\j_g\}=-\2Gp^2+\Psi p(p_\l-\frac{i}{2}\l)+2\Psi\p_c\g+\p_c
p_b+\p_\b ip_\g
\ee
Eliminating $\p_c$, $p_b$, $p_\g$ and $\p_b$ yields
\be
\dot{c}=p_b, \ \ \dot{\g}=-i\p_\b
\ee
and inserting these result back into the action yields
\bea
L \rav\dot{\v} p +\.{\l}\p_\l+\.Gp_G+\.\Psi \p_\Psi \nonumber \\
\ram \2 Gp^2+ \Psi p(\pi_\l-\frac{i}{2}\l)+\.b (\.c+2\g\Psi)+\.\b
(i\.\g)
\eea
The first line contains the kinetic terms and the gauge fixing terms
$\.G=\.\Psi=0$, the second line contains the two first class
constraints and the ghost actions. (The gauge fixing fermion with
$G$ and $\j$ in the Hamiltonian approach has led to gauge fixing
term with $\.G$ and $\.\j$ in the Lagrangian approach. One can
also get $G$ and $\j$ in the Lagrangian approach if one takes
singular limits).

The BC which follow from the field equations are
\bea
&& \v=0 \ \mbox{ or } \ p=0, \ \l=0 \ \mbox{ or } \ \p_\l=0, \ G=0
\
\mbox{ or } \ p_G=0, \nonumber \\ && \Psi =0 \ \mbox{ or } \ \p_\Psi=0,
 b=0 \ \mbox{ or } \ \.c+2\g\Psi =0, \ \.b=0 \ \mbox{ or } \ c=0,\\
&&\b=0 \ \mbox{ or } \ \.\g=0, \ \.\b=0 \ \mbox{ or } \ \g=0
\nonumber
\eea

The BRST transformation rules for this model read
\begin{equation}\label{delb}
\begin{array}{l}
\delta_B\varphi = 
c p \Lambda - i (\pi_{\lambda} -\frac{i}{2}\lambda)\gamma\Lambda, \\
\delta_{B}\lambda = i\gamma p \Lambda, \\
\delta_{B}G = p_b \Lambda,\\
\delta_{B} c = i\gamma \gamma \Lambda,\\
\delta_{B}\gamma=0,\\
\d_B\pi_c =-\frac12 p^2 \Lambda \\
\d_B p_G = \d_B\pi_\Psi = \d_Bp_b=\d_B\pi_\b = 0,
\end{array}
\begin{array}{l}
\delta_{B}p=0,\\
 \delta_{B}\pi_{\lambda} =\frac12 \g p \Lambda , \\
\delta_B{\Psi}=-\pi_{\beta} \Lambda, \\
\delta_{B}b=-{\Lambda} p_G, \\
\delta_{B}\beta=\pi_\psi\Lambda, \\
\d_B p_\g =2i\pi_c\g\Lambda +ip(\pi_\l-\frac{i}{2}\l).
\end{array}
\end{equation}
Then the boundary terms which should vanish if BRST symmetry is to
be exact, are given by
\be
cp^2-i(\p_\l-\frac{i}{2}\l)\g p- i\g p\p_\l+p_bp_G+\p_\b\p_\j
\ee

There are many solutions.
One consistent set of BC for BRST symmetry is
\bea
&& c=0, \ \g=0 \ \ (\mbox{corresponding to $\hat\xi=0$ and $\e=0$})\\
&& b=0, \ \b=0 \ \ (\mbox{since $\.c+2\g\j=0$ and $\.\g=0$ are
ruled out})\\
&& p_G=0, \ \p_\Psi =0 \ \ (\mbox{these are the BRST auxiliary fields})\\
&& \v=0 \ \mbox{ or } \ p=0 \ (\mbox{Dirichlet or Neumann})\\
&& \l=0 \ \mbox{ or } \ \p_\l=0
\eea
In the Hamiltonian formalism, $\p_\l+\frac{i}{2}\l=0$ is a second
class constraint, so $\p_\l$ transforms like $-\frac{i}{2}\l$, and
one may therefore replace $\p_\l$ by $-\frac{i}{2}\l$. Then there
is only one BC on $\l$, namely $\l=0$.

Consistency requires now that also the BRST variation of these
invariants vanish. This is the case.

The BRST charge in (\ref{pdv4}) vanishes provided
\be
c=0, \ \g=0, \ p_G=0, \ \p_\Psi =0
\ee
This only a subset of our consistent set of BC.

The conclusion is that requiring the BRST charge to vanish at
initial or final times leads only to a subset of all BC needed for
consistency as we have defined it. 
Probably, in addition to the vanishing BRST charge, one should also
claim that the symplectic structure is well defined (see
\cite{Papadimitriou:2005ii} and earlier papers \cite{Wald}).
A general discussion of BC in the Hamiltonian formulation of
gravity theories can be found in \cite{ReTe}.

One can also apply the framework of the Hamiltonian approach to
boundaries in a space-like direction. In this case the BC become
Hamiltonian constraints and modify the Dirac brackets between
boundary values of the fields. A general framework for this
procedure was developed in the papers \cite{SolBer} where one can
also find further references. More recently this approach was
applied to the Dirichlet branes \cite{Dbranes}. 
%%%%%%%
\section{$N=1$ supergravity in $2+1$ dimensions}\label{sec21}
In this section we present an example of a complete set of
consistent boundary conditions for a full nonlinear supergravity
model. As model we choose supergravity in $2+1$ dimensions
which is a bit simpler than supergravity in $3+1$ dimensions.
(Note that in $1+1$ dimensions no gauge action for supergravity
exists).
The Lagrangian of simple ($N=1$) supergravity in 3-dimensional
Minkowski space reads
\begin{equation}
\mathcal{L}= -\frac e{2\kappa^2} {R_{\mu\nu}}^{mn}e_m^\nu e_n^\mu 
 +\frac 12 \bar\psi_\mu D_\rho \psi_\sigma \varepsilon^{\mu\rho\sigma}
-\frac e2 S^2 \,.\label{3suact}
\end{equation}
where the real scalar $S$ is an auxiliary field and $D_\rho \psi_\sigma =
\partial_\rho \psi_\sigma +\frac 14 \omega_\rho^{nm}\gamma_{nm}$. Hence
\begin{eqnarray}
&&[D_\mu,D_\nu]=\frac 14 \gamma_{mn}{R_{\mu\nu}}^{mn} \,,\nonumber\\
&&{R_{\mu\nu}}^{mn}=\partial_\mu {\omega_\nu}^{mn}
-\partial_\nu {\omega_\mu}^{mn} +{\omega_{\mu\ \ k}^{\ m}}\, {\omega_\nu}^{kn}
-{{\omega_\nu}^m}_k\, {\omega_\mu}^{kn}.
\label{use2}
\end{eqnarray}
Simple counting of the number of field components minus the number of
local symmetries explains why there is only one scalar auxiliary field:
$9(e_\mu^m)-3(\mbox{Einstein})-3(\mbox{Lorentz})=3$ bosonic components,
$6(\psi_\mu)-2(\mbox{local susy}) =4$ fermionic components.
We shall use the following definition and
identities
\begin{equation}
\varepsilon^{012}=-\varepsilon_{012}=1\,,\quad 
e\gamma^{\mu\nu\rho}=-\varepsilon^{\mu\nu\rho},
\quad \varepsilon^{\mu\rho\sigma}\gamma_\mu =-\gamma^{\rho\sigma}\,,\quad
\gamma^{\mu\nu}\gamma_\nu =2\gamma^\mu \,.\label{use3}
\end{equation}
We use the $1.5$ order formalism, meaning that $\omega_\mu^{mn}$ is determined
by solving its own algebraic field equations \cite{Nieuwenhuizen:1981ae}. 
As a consequence one never needs
to vary  (the $e_\mu^m$ or $\psi_\mu$ in) 
${\omega_\mu}^{mn}$ when one varies the bulk
action, but only the $e_\mu^m$ and $\psi_\mu$ which are explicitly shown
in (\ref{3suact}).
However, varying the $e_\mu^m$ and $\psi_\mu$ term in ${\omega_\mu}^{mn}$
leads to boundary terms which we shall analyze.
The local susy transformations read
\begin{eqnarray}
&&\delta e_\mu^m =\frac {\kappa}2 \bar \epsilon \gamma^m \psi_\mu,\qquad
\delta \psi_\mu =\frac 1{\kappa} D_\mu \epsilon 
+\frac 1{2\sqrt{2}} \gamma_\mu S\epsilon 
\nonumber\\
&&\delta S=\frac 1{2\sqrt{2}} 
\bar \epsilon \gamma^{\mu\nu} (D_\mu \psi_\nu)^{\rm cov}
\,,\label{susytrafo}
\end{eqnarray}
where 
\begin{equation}
(D_\mu \psi_\nu )^{\rm cov}=D_\mu\psi_\nu 
-\frac 1{2\sqrt{2}} \kappa S\gamma_\nu \psi_\mu 
\label{supercov}
\end{equation}
is the supercovariant field strength of the gravitino. 
(The local susy variation of (\ref{supercov}) contains no term
with $\partial_\mu \epsilon$).
The spin connection is also supercovariant and given by
\cite{Nieuwenhuizen:1981ae}
\begin{equation}
\omega_\mu^{\ \, mn}=\omega_\mu^{\ \, mn}(e) +
\frac {\kappa^2}4 \left(
\bar \psi_\mu \gamma^m \psi^n -\bar\psi_\mu \gamma^n \psi^m +
\bar\psi^m \gamma_\mu \psi^n \right) \label{spincon}
\end{equation}
where
\begin{eqnarray}
&&\omega_{\mu mn}(e)=\frac 12 e_m^\nu \left( \partial_\mu e_{n\nu} -
\partial_\nu e_{n\mu} \right) \nonumber\\
&&\qquad -\frac 12 e_n^\nu \left( \partial_\mu e_{m\nu} -
\partial_\nu e_{m\mu} \right) -\frac 12 e_m^\rho e_n^\sigma
\left( \partial_\rho e_{p\sigma} -\partial_\sigma e_{p\rho}\right)
e_\mu^p .\label{cone}
\end{eqnarray}
Again the local susy variation of (\ref{spincon}) contains no terms
with a derivative of $\epsilon$.

We choose Gaussian coordinates, so that
$x^\nor$ is the arc length along geodesics normal to the boundary,
and 
$x^\ti$, $x^\tj$ are coordinates in the surface. Then
\begin{equation}
g_{\nor\nor}=1, \qquad g_{\nor\tj}=g^{\nor\tk}=0 .\label{metric}
\end{equation} 
The normal vector is given by $n^\mu = g^{t\mu}/(g^{tt})^{1/2}$ in a general
coordinate system, but in Gaussian coordinates
\begin{equation}
n^\nu =n_\nu = \delta_\nu^\nor \,.\label{normal}
\end{equation}
We can use this equation to extend $n$ to a vicinity of the boundary.
We also choose the Lorentz indices in such way that the vielbein fields
are block-diagonal
\begin{equation}
e_\nor^N=1, \quad e_\nor^a = 0, \quad e_\tj^N=0\,,
\quad e_\tj^a \ \mbox{arbitrary}\label{vielbein}
\end{equation}
where $N$ and $a$ are flat indices corresponding to the curved indices
$\nor$ and $\tj$.
The use of this special coordinate system considerably simplifies
calculations (see Appendix \ref{ext-app} for technical details). We 
stress that we do not suppose that the variations of the fields also
satisfy (\ref{metric}) and (\ref{vielbein}). The vielbein defined
by (\ref{vielbein}) is not invariant under the diffeomorphism and
Lorentz transformations on the boundary. Therefore, one has to be
very careful when using the coordinate system defined above.
For example, symmetry variations of normal components of the fields
contain both the normal components of usual symmetry variations
{\em and} also the terms with symmetry variations of the normal
vector itself.

First we consider the model without the York-Gibbons-Hawking boundary
term.
From the Euler-Lagrange variational equations one finds the following
BC
\begin{equation}
\left( \delta_e \omega_\tj^{mn} \right) e_m^\tj e_n^\nor \oB =0 \label{1423}
\end{equation}
and 
\begin{equation}
\left[
-\frac e{\kappa^2} \left( \delta_\psi \omega_\tj^{mn} \right) e_m^\tj e_n^\nor
-\frac 12 \varepsilon^{\ti\tj\nor} \bar\psi_\ti \delta \psi_\tj \right]\oB =0.
\label{1426}
\end{equation}
The last term in (\ref{1426}) vanishes whenever $P_+\psi_\tj\oB=0$ or
$P_-\psi_\tj\oB=0$. Then both the variation 
$\delta \omega_\tj^{mn}  e_m^\tj e_n^\nor$
induced by the
variation of the vielbein, and also the variation 
$\delta \omega_\tj^{mn} e_m^\tj e_n^\nor$ induced by
the variation of the gravitino, must vanish at the boundary. 
Hence {\em any} variation
of this component of the spin connection vanishes.
One has to resolve the BC on the components of the spin connection
to obtain corresponding BC for the vielbein and for the gravitino.
We postpone this task for a while and continue to analyze the BC
on the spin connection.

The boundary terms due to a local susy variation read
\begin{equation}
n_\nu \left[ \frac 1{2\kappa}\bar\epsilon D_\ti\psi_\tj 
\varepsilon^{\ti\tj \nu}
-\frac 1{4\sqrt{2}} 
\bar\psi_\ti \gamma_\tj \epsilon S\varepsilon^{\nu\ti\tj}
-\frac e{\kappa^2}e_m^\rho e_n^\nu \delta_\epsilon \omega_\rho^{mn}\right]
\label{bvar}
\end{equation}
where $n$ is an outward pointing unit vector. The first variation is due
to partial integration of 
$\delta \bar\psi_\mu =\partial_\mu\bar\epsilon +\dots$ which is 
needed to produce a curvature (this curvature subsequently cancels against
another curvature which is obtained by varying the explicit vielbeins
in the Einstein-Hilbert action). The second variation is due to 
partial integration of $\partial_\rho \delta \psi_\sigma \sim
\partial_\rho (\gamma_\sigma S \epsilon )$, and the last term is due
to varying all fields in the spin connection.

We are still considering the case
without Gibbons-Hawking boundary term. In that case we know from the 
BC (\ref{set2-4}) of the
linearized theory that $\bar\psi_\ti \gamma_\tj \epsilon$ is nonvanishing,
hence the auxiliary field must satisfy the BC
\begin{equation}
S\oB =0.\label{bconS}
\end{equation}
As we explained before the last term in (\ref{bvar}) vanishes. The first 
term can be solved by inserting $P_++P_-=I$ into it and using that 
$P_-\psi_\tj\oB =0$ and $\bar\epsilon P_+\oB=0$. In Gaussian coordinates
$P_\pm$ commute with the ordinary derivative $\partial_\ti$, 
so that one only finds the BC
\begin{equation}
\omega_\ti^{aN}\oB =0.\label{omegabc}
\end{equation}
If one now uses the BC from the field equations according to which 
both $\delta_e\omega_\tj^{aN}\oB =0$ and 
$\delta_\psi\omega_\tj^{aN}\oB =0$, one sees that the BC (\ref{omegabc})
is consistent: {\em any} variation of this BC also vanishes.
Next we compare (\ref{omegabc}) with (\ref{baroi}) to see that the
following BC holds
\begin{equation}
K^{\ti\tj}\oB =0\,.\label{KijoB}
\end{equation}
The gravitino part of the the condition (\ref{omegabc}) yields
for the contractions with $e_a^\ti$
\begin{equation}
0=(e^\ti_a \omega_\ti^{aN}(\psi ))\oB =
\left[ \frac {\kappa^2}2 \bar \psi_\tj \gamma^\tj \psi^N -
\frac {\kappa^2}4 \bar \psi_\tj \gamma^n \psi^\tj \right]\oB \,.
\label{1737}
\end{equation}
The last term in this equation vanishes, while the first term
yields \\ $\bar \psi_\tj \gamma^\tj P_+ \psi^N\oB =0$, or
\begin{equation}
P_+\psi_N \oB =0 .\label{psinor} \end{equation}
Then all of (\ref{omegabc}) vanishes.

Finally we resolve the BC on the spin connection obtained above and
 close the orbit of the BC (\ref{KijoB}) and (\ref{psinor}) 
to obtain
the following set of BS
\begin{eqnarray}
&&\partial_\nor e_\tj^a \oB =0\,,\qquad \delta e_\nor^a \oB =0\,,\qquad
\delta e_\tj^N\oB =0 \label{ebc}\\
&&P_-\psi_\tj\oB =0,\qquad P_+\psi_\nor \oB =0,\label{psiDbc}\\
&&P_+\partial_\nor \psi_\tj\oB =0,
\qquad P_-\partial_\nor \psi_\nor \oB =0 ,\label{psiNbc}\\
&&S\oB =0 \,.\label{Sbc}
\end{eqnarray}
The BC on the 
parameters are as follows
\begin{eqnarray}
&&\xi^\nor \oB =0,\qquad \partial_\nor \xi^\tj \oB =0, \label{xibc}\\
&&P_-\epsilon \oB =0,\qquad P_+\partial_\nor \epsilon =0,\label{epsbc}\\
&&\lambda^{aN}\oB =0,\qquad \partial_\nor \lambda^{ab}\oB =0. \label{lambc}
\end{eqnarray}
Here $\lambda^{mn}$ are parameters of the Lorentz transformations.
Symmetry considerations do not require any restrictions on $e_\nor^N$.
However, if one wishes to impose a BC on $e_\nor^N$, this condition 
must be Neumann
\begin{equation}
\partial_\nor e_\nor^N \oB =0,\label{dnnN}
\end{equation}
since the Dirichlet condition violates susy.

By using the formulae from Appendix \ref{appB} one can easily check that
the boundary conditions (\ref{ebc}) - (\ref{Sbc}) are closed
under the action of all local symmetries provided the transformation
parameters satisfy (\ref{xibc}) - (\ref{lambc}). The BC
(\ref{omegabc}) and (\ref{KijoB}) are satisfied as well.

Let us now turn to the case when the Gibbons-Hawking boundary term 
is added to the action, and when one chooses therefore the Gibbons-Hawking
BC for the gravity fluctuations
\begin{equation}
\delta g_{\ti\tj} \oB =0 \,.\label{anotherGH}
\end{equation}
This case was studied by Luckock and Moss \cite{Luckock:1989jr}. We shall
use many results from that paper. 
%The authors of \cite{Luckock:1989jr}
%found a set of boundary fields which transform into each other under the
%susy transformations on the boundary. 
No locally invariant set of
BC was presented in \cite{Luckock:1989jr}. By modifying 
slightly the analysis
of that paper we show that such a set does not exist.

Obviously, the BC on the spin connection (\ref{omegabc}) are not satisfied
in the presence of the York-Gibbons-Hawking term.
Also in the presence of this
term the diffeomorphism invariance of the action implies the BC
\begin{equation}
\xi^\nor \oB =0 \label{1stdif}
\end{equation}
on the parameter $\xi^\mu$. Consistency requires that also
\begin{equation}
0=\delta_\xi g_{\ti\tj} \oB = \left( \xi_{\ti :\tj} +
\xi_{\tj:\ti} -2\Gamma_{\ti\tj}^\nor \xi_\nor \right)\oB ,\label{1826}
\end{equation}
where the colon denotes covariant differentiation with the Christoffel
symbol constructed from the metric of the boundary. Together with
the BC (\ref{1stdif}) the equation (\ref{1826}) tells us that
$\xi_\tj$ is a Killing vector on the boundary. There is at most a finite
number of Killing vectors which generate rigid symmetries. 
In this section we are interested
in local symmetries, so that we assume for simplicity that there are
no Killing vectors on $\partial\mathcal{M}$. Consequently, we have
the following BC
\begin{equation}
\xi_\ti \oB =0 \,.\label{2nddif}
\end{equation}
Closure of the susy algebra then requires 
$\bar \epsilon_2 \gamma_\mu \epsilon_1\oB =0$ for all indices $\mu$.
This condition yields
\begin{equation}
\epsilon \oB =0\,. \label{alleps}
\end{equation}

If one wishes to couple this system to a spin 1 field, our linearized
analysis shows (see eq.\ (\ref{npsin}) that the gravitino must satisfy
\begin{equation}
P_-\psi_\nor \oB =0 \,.\label{1900}
\end{equation}
Even if no spin 1 fields are present, a more tedious analysis of
hermiticity properties of the fluctuation operators \cite{Luckock:1989jr}
also requires the same condition (\ref{1900}) of the 
gravitino\footnote{This condition follows from the equations (6.48),
(6.49) and the definition (6.20) of  \cite{Luckock:1989jr}. Note that
compared to that paper we have interchanged the role of $P_+$ and
$P_-$.}. Consistency now requires
\begin{equation}
0=\delta_\epsilon P_-\psi_\nor \oB = P_-\partial_\nor \epsilon \oB \,,
\label{1907}
\end{equation}
where we used (\ref{alleps}).
Therefore, one has {\em both} Dirichlet and Neumann BC on $P_-\epsilon$.
This clearly excludes local susy transformations on the boundary.  

We must stress that this conclusion is valid for local BC only.
If one allows for nonlocal BC, one can easily resolve the contradiction
we have found above. However, it still remains an open question
whether one can find a closed consistent orbit of BC with nonlocal
BC.
%%%%%
\section{Conclusions and comments}\label{sec-conc}
In this article we have determined the complete consistent set 
(``the orbit'') of BC for supergravity models, which maintains local
susy even at the boundaries. Violation of local susy by boundaries
may not be fatal, it may perhaps even be welcome, but we have
studied when local susy remains unbroken. We have worked completely at the 
classical level; at the quantum level, boundary term may also be needed
to remove infrared divergences \cite{Larsen:2003pf}. The renormalization
group flow affects the boundary conditions \cite{Schalm:2004qk}.

Our main result is that local susy of the BC in supergravity requires
vanishing extrinsic curvature $K^{\tj\ti}\oB =0$. The surfaces with
zero extrinsic curvature are called totally geodesic. Such surfaces
contain geodesics connecting any two points belonging to them.
It is interesting to note that totally geodesic surfaces are also
minimal, i.e. they are solutions of the classical equations of
the bosonic $p$-branes.

We considered massless fields in the text. For massive fields
there are differences. Consider the massive spin 0 - spin $1/2$
system with $\delta_\epsilon \lambda = \delta_\epsilon (m=0)\lambda
+m(S+iP)\epsilon$. For the BC
 $P_-\epsilon =0$, $P_+\lambda \oB =\partial_\nor S\oB=P\oB=0$.
  one finds from $P_\pm \delta_\epsilon \lambda =0$
the BC
\begin{equation}
(\partial_\nor \pm m) S=0.\label{massbc}
\end{equation}
since $P_\pm\gamma^\nor = \pm P_\pm$ if $(\gamma^\nor )^2=1$.
Similarly for interacting theories there are differences
(see section \ref{sec-kink}). 

In the case of pure gravity, gauge or BRST invariant BC
for the graviton were constructed in \cite{gravBC,Moss:1996ip}
(see \cite{EKPbook} for an overview). It is interesting to note
that the BC obtained in \cite{gravBC} either contain tangential
derivatives of the fields on the boundary, or admit noncovariant
gauges only. The sets BC of \cite{Moss:1996ip}, which do not
depend on tangential derivatives were obtained with some restrictions
on the extrinsic curvature of the boundary. In the presence of 
tangential derivatives in BC, quantum loop calculations become
extremely difficult.   

Local BC for supergravity were considered in 
\cite{D'Eath:1984sh,Luckock:1989jr,Esposito:1996kv,EKPbook}.
All these papers started with the Gibbons-Hawking condition
(\ref{dgij}) on the gravity fluctuations. Therefore, it was not
possible to obtain a fully locally supersymmetric set of BC. 
There exists a great variety of types of BC which are being
used in quantum field theory (see reviews 
\cite{Vassilevich:2003xt,Vassilevich:2004id}).
Here we did not consider nonlocal BC of the Atiyah-Patodi-Singer type
or other exotic conditions. Nonlocal BC in supergravity were
studied in \cite{Eath:1991sz,EKPbook,Esposito:1996kv}, but no
locally susy orbit of such BC was found. In may applications it is
desirable to have the Gibbons-Hawking BC for the gravity fluctuations,
or at least the York-Gibbons-Hawking boundary term in the action
(see, e.g., \cite{Liu:1998bu,Dvali:2003zh}). Since our no-go result
is valid for local BC only, one should probably reconsider the nonlocal
option again. 

Globally supersymmetric asymptotic conditions were constructed by 
Breitenlohner and Freedman \cite{BF} who considered gauged supergravity
in AdS. They obtained two sets of the asymptotic conditions which 
correspond\footnote{One has to note that asymptotic conditions
are not the same as BC. Nevertheless, one can map one into the other
by identifying the fields which vanish fast at the infinity with the
fields which satisfy Dirichlet BC. We also like to mention here 
related works \cite{HennTeit} where boundary terms in the AdS space
and their properties with respect to rigid symmetry transformations
were analysed. Various choice of BC in AdS were discussed
recently in \cite{AdSBC,Papadimitriou:2005ii}.} to our sets (\ref{set1-4})
and (\ref{set2-4}). Later Hawking \cite{Haw83} suggested an additional
requirement to choose between these two sets. He required that the 
space-time approaches anti-de Sitter space sufficiently fast at infinity
that the asymptotic group of motion of the space-time is the AdS
group $O(3,2)$. In this case there exist asymptotically supercovariant 
constant spinors which generate asymptotic global susy transformations.
If one now demands that the space-time remains asymptotically AdS,
one is led to the asymptotic conditions which correspond to the
second set (\ref{set2-4}). This is precisely the set of BC which
was selected in section \ref{sec21} as preserving {\em local}
susy on the boundary. Therefore, certain problem with supersymmetries in
the first set (\ref{set1-4}) were noted long ago \cite{Haw83} though
not for BC but rather for asymptotic conditions on AdS. This
is particularly remarkable given large interest to supergravities
and AdS in general and to the asymptotic conditions on graviton 
in particular \cite{Liu:1998bu}.

Matching conditions on a brane which restrict the extrinsic curvature were
studied in supergravity 
only recently by Moss \cite{Moss:2004ck}. That paper, however, did not
analyse the closure of the set of matching conditions under all symmetry
transformations.

\section*{Note added in proof}
If one removes some of our requirements, e.g. if one does not
require the consistency of the boundary conditions with the
equations of motion, then the York-Gibbons-Hawking action
can be made compatible with local susy \cite{44}. We are grateful
to Dmitry Belyaev for explaining this point to us. We also
like to mention the work \cite{45} where boundary terms for the
Lovelock gravity were studied. For a
discussion of boundary terms which remove second order derivatives
from
all fields in the Hilbert-Einstein action, see \cite{46}.

\section*{Acknowledgements}
We are very grateful to Ian Moss for many email exchanges and details on
calculations with extrinsic curvatures. Giampiero Esposito also gave us useful
information. Closer to home, we had helpful discussions with Martin Rocek and
Warren Siegel. DV thanks the C.~N.~Yang ITP for the support of his visit
to Stony Brook. He was also supported by the DFG project BO 1112/12-2. 
%%%%%5
\appendix
\section{Extrinsic curvature}\label{ext-app}
The extrinsic curvature is defined by the relation
$K_{\mu\nu}=(g_{\mu\rho}\mp n_\mu n_\rho )(g_{\nu\sigma} \mp n_\nu n_\sigma)
D^\rho n^\sigma$. The induced metrics $g_{\mu\nu}\mp n_\mu n_\nu$
yield projection operators $g^{\mu\rho}(g_{\rho\nu}\mp n_\rho n_\nu)=
\delta^\mu_{\ nu}\mp n^\mu n_\nu$ if $g^{\mu\nu}n_\mu n_\mu =\pm 1$.
We continue with the upper sign.
Since $\partial_\rho (n^\mu n_\mu) =0=2n_\mu D_\rho n^\mu$,
it can be simplified to
\begin{equation}
K_{\mu\nu} = n_{\nu;\mu} -n_{\mu} n^{\rho} n_{\nu;\rho} \,.\label{extdef}
\end{equation}
In Gaussian coordinates $n_\nu = (1,0,\dots 0)$, $g_{\nor\nor}=g^{\nor\nor}=1$,
and $g_{\nor\ti}=g^{\nor\ti}=0$, hence
\begin{equation}
K_{\ti\tj}= -\Gamma_{\ti\tj}^\nor 
=\frac 12 \partial_\nor g_{\ti\tj}; \qquad 
K^{\ti\tj}=-\frac 12 \partial_n g^{\ti\tj}
 \,,\label{extcursp}
\end{equation}
and all other components of $K_{\mu\nu}$ vanish. 

Next we give 
several useful relations between variations of the metric
and the normal vector. We suppose that before the variation the
``background'' metric (denoted by $\bar g_{\mu\nu}$) is Gaussian.
The full metric $g_{\mu\nu}=\bar g_{\mu\nu}+
\delta g_{\mu\nu}$ is not Gaussian, of course, but the varied normal
$n^\mu =\bar n^\mu +\delta n^\mu$ is still perpendicular to the surface,
$n_\ti =0$, and normalized to unity, $n^\mu g_{\mu\nu} n^\nu =1$. 
The variation of $g^{\mu\nu}$
can be expressed in terms of $\delta g_{\mu\nu}$ as follows
\begin{equation}
\delta g^{\ti\tj}=-\bar g^{\ti\tk} \bar g^{\tj\tl} \delta g_{\tk\tl}\,,\quad
\delta g^{\ti\nor}=-\bar g^{\ti \tk} \delta g_{\tk\nor}\,,\quad
\delta g^{\nor\nor}=-\delta g_{\nor\nor}\,.\label{gdown}
\end{equation}
Under arbitrary variations of the metric the normal varies as follows
\begin{eqnarray}
&&\delta n^\nor =-\frac 12 \delta g_{\nor\nor}\,,
\qquad \delta n^\tj =-\bar g^{\tj\tk}\delta g_{\nor\tk}\,,\nonumber\\
&&\delta n_\nor =\frac 12 \delta g_{\nor\nor}\,,\qquad \delta n_\tj =0\,.
\label{varn}
\end{eqnarray}
It is straightforward to prove that the variation of extrinsic curvature
reads
\begin{eqnarray}
&&\delta K_{\nor\nor}=0,\nonumber \\
&&\delta K_{\nor\tj} = \bar K_\tj^\tk \delta g_{\nor\tk},\nonumber\\
&&\delta K_{\ti\tj} = -\frac 12 \bar K_{\ti\tj} \delta g_{\nor\nor}
-\frac 12 \left[ (\delta g_{\ti\nor})_{:\tj} + (\delta g_{\tj\nor})_{:\ti}
-\partial_{\nor} \delta g_{\ti\tj} \right] .\label{varK}
\end{eqnarray}
The colon denotes covariant differentiation with the Christoffel symbol
defined by the metric $\bar g_{\tj\tk}$.

The equations (\ref{gdown}) - (\ref{varK}) do not use any boundary conditions
and, therefore, can be differentiated with respect to $x^\nor$. Note, that we
have extended the normal vector and, consequently, the extrinsic curvature 
to outside the boundary. In our coordinate system 
\begin{equation}
\bar \Gamma_{\nor\nor}^\nor = \bar \Gamma_{\nor\nor}^\tk
=\bar \Gamma_{\nor\tk}^\nor = 0\,,\qquad
\bar \Gamma_{\tj\tk}^\nor =-\bar K_{\tj\tk}\,,\qquad
\bar \Gamma_{\nor\tk}^\tj =\bar K^\tj_\tk \label{Gamma-id}
\end{equation}

\section{Torsion}\label{appB}
Consider the bosonic part of the spin-connection. It can be defined
through the vielbein equation
\begin{equation}
D_\mu e_\nu^m = \partial_\mu e_\nu^m -\Gamma_{\mu\nu}^\rho e_\rho^m
+{{\omega_\mu}^m}_p(e) e_\nu^p =0 ,.\label{vieleq} 
\end{equation}
From this equation we have the following components of the connection
$\bar \omega_\mu (\bar e)$ in the adapted coordinate system
(\ref{metric}) - (\ref{vielbein}):
\begin{eqnarray}
&&{\bar\omega_\nor}^{\ \, aN}=0,\qquad 
{\bar\omega_\nor}^{\ \, ab} =-\bar e^{\tj b}
\partial_\nor \bar e_\tj^a -\bar K_{\tj\ti}\bar e^{\tj a} \bar e^{\ti b},
\nonumber\\ %\label{baron}\\
&&{\bar\omega_\ti}^{\ \, aN}=\bar K_{\ti\tj} 
\bar e^{\tj a},\qquad 
{\bar \omega_\ti}^{\ \, ab}=-\bar e^{\tj b}\partial_\ti \bar e_\tj^a
+\bar \Gamma_{\ti\tj}^\tk \bar e_\tk^a \bar e^{\tj b} .
\label{baroi}
\end{eqnarray}
This formulae imply that on the boundary all components of 
$\bar \omega_\mu (\bar e)$  
except for $\bar \omega_\tj^{ab}$ vanish for the BC (\ref{KijoB}).

Next we study the gravitino part of the connection ${\omega_\mu}^{mp}(\psi)$.
One can prove that our boundary conditions (\ref{psiDbc}) yield
\begin{equation}
{\omega_\nor}^{ab}(\psi)\oB =0,\qquad {\omega_\tj}^{aN}(\psi)\oB =0.
\label{opsib}
\end{equation}
Consider first 
\begin{equation}
{\omega_\nor}^{ab}(\psi) 
=\frac {\kappa^2}4 \left( \bar\psi_\nor \gamma^a \psi^b
-\bar \psi_\nor \gamma^b \psi^a +\bar\psi^a \gamma_\nor \psi^b \right)\,.
\label{omnabps}
\end{equation}
On the boundary, the following identities hold
\begin{eqnarray}
&&\bar\psi_\nor \gamma^a \psi^b = \bar\psi_\nor \gamma^a (P_++P_-)\psi^b =
\bar\psi_\nor \gamma^a P_+\psi^b =\bar\psi_\nor P_- \gamma^a \psi^b =0,
\nonumber\\
&&\bar\psi^a \gamma_\nor \psi^b = \bar\psi^a \gamma_\nor (P_++P_-) \psi^b
=\bar\psi^a \gamma_\nor P_+ \psi^b =\bar\psi^a P_+ \gamma_\nor \psi^b =0.
\end{eqnarray}
The first equality in (\ref{omnabps}) is now obvious, the second one can be
demonstrated in a similar manner.

\end{document}